\documentclass{gGAF2e}
\usepackage{graphicx,natbib,bm,url,color}
\newcommand{\dd}{{\rm d} {}}
\newcommand{\K}{\,{\rm K}}
\newcommand{\s}{\,{\rm s}}
\newcommand{\us}{\,\mu{\rm s}}
\newcommand{\ps}{\,{\rm ps}}

\newcommand{\GHz}{\,{\rm GHz}}
\newcommand{\um}{\,\mu{\rm m}}
\newcommand{\cm}{\,{\rm cm}}
\newcommand{\km}{\,{\rm km}}
\newcommand{\dyn}{\,{\rm dyn}}
\newcommand{\bbar}{\,{\rm bar}}
\newcommand{\Sec}[1]{section~\ref{#1}}
\newcommand{\Fig}[1]{figure~\ref{#1}}
\newcommand{\Tab}[1]{table~\ref{#1}}
\newcommand{\Eq}[1]{equation~(\ref{#1})}
\newcommand{\EQ}{\begin{equation}}
\newcommand{\EN}{\end{equation}}

\newcommand{\bra}[1]{\langle #1\rangle}
\def\nab{{\bm{\nabla}}}
\def\ttau{{\bm{\tau}}}

\def\JJ{{\bm{J}}}
\def\UU{{\bm{U}}}
\def\VV{{\bm{V}}}
\def\qq{{\bm{q}}}

\begin{document}

\jvol{00} \jnum{00} \jyear{2019} 

\markboth{\rm C. QIAN ET AL.}{\rm GEOPHYSICAL \& ASTROPHYSICAL FLUID DYNAMICS}


\title{Convergence properties of detonation simulations}

\author{Chengeng Qian$^{\rm a}$,
Cheng Wang$^{\rm a}$,
JianNan Liu$^{\rm b}$,
Axel Brandenburg$^{\rm c,d,e}$$^{\ast}$\thanks{$^\ast$Email: brandenb@nordita.org},\\
Nils E.L. Haugen$^{\rm f}$ and
Mikhael A. Liberman$^{\rm c}$\vspace{6pt}
\\
$^{\rm a}$State Key Laboratory of Explosion Science and Technology, Beijing Institute of Technology,\\Beijing, 100081, China\\
$^{\rm b}$College of Mining Engineering, Taiyuan University of Technology, Taiyuan 030024, China\\
$^{\rm c}$Nordita, KTH Royal Institute of Technology and Stockholm University,\\ Roslagstullsbacken 23, SE-10691 Stockholm, Sweden\\
$^{\rm d}$Department of Astronomy, Stockholm University, SE-10691 Stockholm, Sweden\\
$^{\rm e}$JILA and Laboratory for Atmospheric and Space Physics, University of Colorado,\\ Boulder, CO 80303, USA\\
$^{\rm f}$SINTEF Energy Research, 7465 Trondheim, Norway\\
\vspace{6pt}\received{\it \today,~ $ $Revision: 1.112 $ $} }

\maketitle

\begin{abstract}
We present a high-resolution convergence study of detonation initiated
by a temperature gradient in a stoichiometric hydrogen--oxygen mixture
using the {\sc Pencil Code} and compare with a code that employs
a fifth order weighted essentially non-oscillating (WENO) scheme.
With Mach numbers reaching $10$--$30$, a certain amount of shock viscosity
is needed in the {\sc Pencil Code} to remove or reduce numerical pressure
oscillations on the grid scale at the position of the shock.
Detonation is found to occur for intermediate values of the shock viscosity
parameter.
At fixed values of this parameter, the numerical error associated with
those small wiggles in the pressure profile is found to decrease with
decreasing mesh width $\delta x$ like $\delta x^{-1.4}$
down to $\delta x=0.2\um$.
With the WENO scheme, solutions are smooth at $\delta x=10\um$, but
no detonation is obtained for $\delta x=5\um$.
This is argued to be an artifact of a decoupling between pressure
and reaction fronts.
\end{abstract}

\section{Introduction}

Detonation can be produced by the coupling of a spontaneous reaction wave,
which propagates along an initial temperature gradient, with a pressure
wave \citep{Zeldovich+70,Z1980}.
This process is governed by the time-dependent compressible reactive
Navier-Stokes equations.
Its direct numerical simulation (DNS) is an intricate problem that is
of fundamental importance for understanding the ignition of different
combustion modes caused by a transient thermal energy deposition localised
in a finite volume of reactive gas \citep{liberman2012regimes}.
High resolution methods are necessary to resolve the broad
range of length scales.
It is also well-known that problems involving strong shocks, such as
in the final stage of the deflagration to detonation transition
(DDT), require the use of shock-capturing techniques to eliminate or
reduce spurious oscillations near discontinuities.
One of the widely used approaches is the use of weighted essentially
non-oscillating (WENO)
finite differences \citep{JS1996}, which is an improvement upon the essentially
non-oscillating (ENO) scheme.
The main idea of the WENO scheme is to use a convex combination of all the
candidate stencils rather than the smoothest candidate stencil to achieve a
higher order accuracy than the ENO scheme, while maintaining the essentially
non-oscillating property near discontinuities.
There are also other methods such as the Total Variation Diminishing (TVD)
method or the Artificial Compression Method (ACM) switch \citep{Lo+07}.

Yet another approach to avoid wiggles in the numerical solution is to
add a shock-capturing viscosity.
However, one must be cautious when using such a shock-capturing viscosity,
since its properties are problem-dependent.
The shock-capturing viscosity will fail to eliminate oscillations if it
is too small.
Since the gaseous combustion process is highly sensitive to the resolution of
the reaction zone, using too large shock-capturing viscosity can lead
to an artificial coupling of the leading pressure wave and the flame front.
Thus, it is essential to determine the proper shock-capturing viscosity when
using the {\sc Pencil Code} to simulate problems involving the onset of detonation.

The test problem examined in this paper is the hot spot problem, which
is a chemically exothermic reactive mixture with a nonuniform
distribution of temperature.
According to the theory developed by \cite{Zeldovich+70} and \cite{Z1980},
the gradient of induction time  associated with temperature (or concentration)
gradients may be ultimately responsible for the detonation initiation.
A similar concept of shock-wave amplification by coherent energy release
(SWACER) was introduced later by \cite {Lee+Moen80}.
The basic idea is that a spontaneous reaction wave can propagate through a
reactive gas mixture if there is a spatial gradient in the chemical induction
time $\tau_{\rm ind}$.
The spontaneous reaction is ignited first at the location of
minimum induction (ignition delay) time $\tau_{\rm ind}$ and then spreads
by spontaneous ignition over neighbouring locations where the temperature is
lower and $\tau_{\rm ind}$ is correspondingly longer.
The velocity of the spontaneous reaction wave is analogous to a phase velocity.
It cannot be smaller than the velocity of deflagration, but is not limited
from above, and depends on the steepness of the temperature gradient and
the temperature derivative of the induction time.
The proposed mechanism of detonation initiated by the temperature
gradient suggests that the formation of an induction time gradient
produces a spatial time sequence of energy release, which then produces
a compression wave that gradually amplifies into a shock wave.
Coupling of the spontaneous reaction and pressure waves can cause shock
wave amplification by coherent energy release and can finally result
in the formation of a detonation wave. This requires a certain synchrony
between the progress of the shock and the sequential release of chemical
energy by successive reactions along of the temperature gradient.

The first numerical demonstration of the formation of a detonation wave by a
temperature gradient was by \cite{Zeldovich+70}.
Although this earliest numerical solution had a low resolution, the
authors demonstrated successfully that sufficiently shallow gradients
produce detonation, while for steeper gradients the reaction wave and
the shock failed to couple together.
In subsequent studies, \cite{Z1988,HE1992,HE1996,KOW1997,B2000}, and
\cite{KR2014} have employed a one-step chemical model to investigate
regimes of detonation ignition by an initial temperature gradient.
However, the one-step model or other simplified chemical models do not
predict correctly the induction time for the combustion process involving
a large set of chain-branching reactions.
\cite{LKI2011, liberman2012regimes} studied different modes of combustion
produced by the initial temperature gradient in stoichiometric
hydrogen--oxygen and hydrogen--air mixtures ignited by a temperature
gradient using detailed chemical models and compared the results with
those obtained with a one-step chemical model.
In particular, it was shown that the minimal
slope of the temperature gradient required for triggering detonation
and other combustion modes obtained in simulations with simplified
chemical models, for example a one-step model, is orders of magnitude
smaller than those obtained in simulations with a detailed chemical model.
\cite{WQLL2018} and \cite{LWQL2018} studied the influence of the chemical
reaction model on detonation ignited by a temperature gradient for
hydrogen-air and methane-air mixtures.
They concluded that the one-step model and other simplified models usually cannot
describe correctly the ignition processes.
Thus, using simplified chemical kinetics for understanding the mechanisms of
DDT must be considered with great caution.
Using one-dimensional Navier-Stokes equations with detailed chemical
kinetics, \cite{Gu2003} extended Zeldovich's temperature gradient 
theory and demonstrated five modes of reaction front propagation from a hot 
spot for hydrogen and syngas mixture at high pressure ($50\bbar$).
They identified the regimes of detonation initiation using two 
dimensionless parameters, namely the ratio of sound speed to reaction front 
velocity and the residence time of the acousic wave in the hot spot 
normalised by the excitation time of the unburned mixture. 
This theory has been employed and extended to investigate the super-knock in gasoline spark ignition
engines \citep{B2009,B2012,RZRA2013,BBP2016,DC2015a,DC2015b}.  

The aim of this paper is to study the convergence of detonations
simulation using the {\sc Pencil Code}.
We also present a comparison with the simulation results of a code that employs a fifth
order WENO scheme.
The paper is organised as follows.
In section~2, the governing equations are presented and the setup of
the hot spot problem is described.
Section~3 presents a convergence study of the pressure profiles obtained
using the {\sc Pencil Code}.
The dependence on the shock viscosity is also investigated in this section.
In section~4, we consider the convergence of the same problem of detonation 
produced by a temperature gradient using the WENO code. 
In \Sec{Concl}, we conclude by summarising our main findings.

\section{The model}

\subsection{The basic equations}

The set of equations for modelling combustion was implemented into
the {\sc Pencil Code} by \citet{BHB11}.
Considering a mixture of $N_{\rm s}$ species undergoing $N_{\rm r}$
reactions, we solve the continuity equation for the total density $\rho$,
\begin{eqnarray}
\frac{{\rm D}\ln \rho}{{\rm D} t} = -\nab {\bm \cdot} \bm{U},
\label{eq:rho}
\end{eqnarray}
the momentum equation for the velocity $\UU$,
\begin{eqnarray}
\frac{{\rm D} {\UU}}{{\rm D} t}= -\frac{1}{\rho}\nab p
+\frac{2}{\rho} {\bm \nabla}{\bm \cdot}\ttau,
\label{eq:UU}
\end{eqnarray}
the energy equation for the temperature $T$,
\begin{eqnarray}
c_v \frac{{\rm D} \ln T}{{\rm D} t}  =
\sum_k^{N_{\rm s}} \frac{{\rm D} Y_k}{{\rm D} t} \left(\frac{R}{W_k}- \frac{h_k}{T} \right)
-\frac{R}{W} \nab {\bm \cdot} {\bm U} +\frac{\ttau:\nab\bm{U}}{\rho T} -\frac{\nab {\bm \cdot} \qq}{\rho T}, \label{eq:energy}
\end{eqnarray}
and the equation for the mass fraction of the $k$th species $Y_k$ in the form
\begin{equation}
\rho \frac{{\rm D} Y_k}{{\rm D} t}  =
-\nab {\bm \cdot} {\JJ_k} + \dot{\omega}_k,
\label{eq:YY}
\end{equation}
where ${\rm D}/{\rm D}t= \partial/\partial t + {\bm U}{\bm \cdot} {\bm \nabla}$
is the advective derivative and
$\tau_{ij}=2\rho\nu{\sf S}_{ij}+\rho\zeta\delta_{ij}\nab{\bm \cdot}\UU$
are the components of the stress tensor with
${\sf S}_{ij}={1\over2}(\partial U_i/\partial x_j+\partial U_j/\partial x_i)
-{1\over3}\delta_{ij}\nabla{\bm \cdot}\bm{U}$ being the components of the
traceless rate-of-strain tensor, $\nu$ is the kinematic viscosity,
$\zeta$ is the bulk viscosity, $\dot{\omega}$ is the reaction rate and
subscript $k$ refers to species number $k$.
The pressure is given by the equation of state,
\begin{equation}
p=\rho T\frac{R}{W}=\rho TR\sum_{k=1}^{N_{\rm s}}\left(\frac{Y_k}{W_k}\right),
\end{equation}
where $R$, $W$, and $W_k$ are the universal gas constant, the mean
molecular weight of the mixture, and the molecular weight of species $k$,
respectively.
The viscosity of species $k$ is given by \cite{ch81} as
\begin{equation}
\mu_k=\frac{5}{16}\frac{\sqrt{\pi k_{\rm B} T m_k}}{\pi\sigma^2_{k}\varOmega^{(2,2)*}_k},
\end{equation}
where $\sigma_k$ is the Lennard-Jones collision diameter, $k_{\rm B}$
is the Boltzmann constant, $m_k$ is the mass of the molecule, and
$\varOmega^{(2,2)*}_k$ is the collision integral \citep[see][]{MR77}.
Then, the viscosity of the mixture, $\mu=\rho\nu_{\rm mix}$, is given by \citep{w50}
\begin{equation}
\mu=\sum_{k=1}^{N_{\rm s}} \left(X_k \mu_k
\left/\sum_{j=1}^{N_{\rm s}} X_j \varPhi_{kj}\right)\right..
\end{equation}
Here, $X_k$ is the mole fraction of species $k$ and $\varPhi_{kj}$ is
given by
\begin{equation}
\varPhi_{kj}=\frac{1}{\sqrt 8} \left(1 + \frac{W_k}{W_j} \right)^{-1/2} \left[
1+\left(\frac{\mu_k}{\mu_j}\right)^{1/2} \left(\frac{W_j}{W_k}\right)^{1/4} \right]^2.
\end{equation}
The heat flux $\qq$ is given by
\begin{equation}
\qq=\sum_{k=1}^{N_{\rm s}}h_k \JJ_{k}-\lambda\nab T.
\end{equation}
Here, $\JJ_k=\rho Y_k\VV_k$ is the diffusive flux.
Fick's law is employed to calculate the diffusion velocity $\VV_k$
as \citep{P2005}
\begin{equation}
{\bm V_k}=-\frac{D_k}{X_k}\nab X_k,
\end{equation}
where the diffusion coefficient for species $k$ is expressed as
\begin{equation}
D_k=\frac{1-Y_k}{\sum_{j\ne k}^{N_{\rm s}}X_j/D_{jk}},
\end{equation}
and the binary diffusion coefficient is given by
\begin{equation}
D_{jk}=\frac{3}{16}\frac{\sqrt{2\pi k_B^3T^3/m_{jk}}}{p\pi\sigma_{jk}^2\varOmega_{jk}^{(1,1)*}},
\end{equation}
where $\varOmega_{jk}^{(1,1)*}$, $\sigma_{jk}$, and $m_{jk}$ are given by
\cite{E2007}.

The thermal conductivity for pure species $k$ is expressed as
\begin{equation}
\lambda_k=\frac{\mu_k}{W_k}\left(f_{\rm trans}{\bm \cdot}
C_{v,\rm trans}+f_{\rm rot}{\bm \cdot} C_{v,\rm rot}+f_{\rm vib}
{\bm \cdot} C_{v,\rm vib}\right),
\end{equation}
and the thermal conductivity of the mixture follows an empirical law.
The specific heat $c_{p,k}$ and specific enthalpy $h_k$ of species $k$ are
calculated by using tabulated polynomials used in rocket science by the
National Aeronautics and Space Administration (NASA) and are known
as NASA polynomials. We use here the coefficients from \cite{K2013}.

The expression for the reaction rate is \citep{P2005}
\begin{equation}
\dot{\omega}_k=W_k\sum_{s=1}^{N_r}\left(\nu_{ks}''-\nu_{ks}'\right)
\left[
k_{{\rm f},s}\prod_{j=1}^{N_{\rm s}}\left(\frac{\rho_j}{W_j}\right)^{\nu_{js}'}-
k_{{\rm r},s}\prod_{j=1}^{N_{\rm s}}\left(\frac{\rho_j}{W_j}\right)^{\nu_{js}''}
\right],
\end{equation}
where $\rho_k$ is the density of species $k$.
Furthermore,
$\nu^{\prime}_{ks}$ and $\nu^{\prime\prime}_{ks}$ are the stoichiometric
coefficients of species $k$ of reaction $s$
on the reactant and product sides, respectively.
Furthermore, $k_{{\rm f},s}$ is the forward rate of reaction $s$,
which is given by
\EQ
k_{{\rm f},s}=B_s T^{\alpha_s}\exp(-E_{s}/RT),
\EN
where $B_s$ is a pre-exponential factor,
$\alpha_s$ is the temperature exponent,
and $E_s$ is the activation energy.
These are all empirical coefficients that
are given by the kinetic mechanism.
The backward rate of reaction $s$ is calculated from the forward rates through 
the equilibrium constant
\begin{equation}
k_{{\rm r},s}=k_{ {\rm f},s}/k_{{\rm c},s},
\end{equation}
where $k_{{\rm c},s}=
(p_0/RT)^{\sum_{k=1}^{N_s}(\nu_{ks}''-\nu_{ks}')}
\exp(\Delta S_s/R-\Delta H_s/RT)$.
Here $p_0=1\bbar$, $\Delta S_s$ and $\Delta H_s$ are entropy and enthalpy
changes for reaction $s$.
The detailed chemical mechanism chosen to simulate the hot spot problem
is the mechanism developed by \cite{K2013}, which includes $N_{\rm r}=19$
reactions and $N_{\rm s}=8$ species.
The induction time of this mechanism, which is one of the important
parameters for the simulation, has been validated by extensive experiments
and simulations at pressure from $1$ to $70\bbar$, over a temperature
range of $914\K$ to $2200\K$.

\subsection{Treatment of shocks in the {\sc Pencil Code} and setup
using the WENO code}
\label{sec:Shocks}

In the {\sc Pencil Code}, the
shock viscosity of \cite{vNR50} is applied as a bulk viscosity,
\begin{equation}
\zeta=C_{\rm shock} \delta x^2 \bra{-\nab{\bm \cdot}\UU}_+,
\label{zeta_shock}
\end{equation}
and is required to eliminate wiggles in the numerical solution.
Here, $\bra{...}_+$ denotes a running five point average over all positive
arguments, corresponding to a compression.

In the WENO code, equations (\ref{eq:rho})--(\ref{eq:YY}) are
solved in the conservation form; see equations (5)--(9) of \cite{WQLL2018}. 
The chemical model for hydrogen-oxygen is the same model as that
developed by \cite{K2013}. 
The one-dimensional simulations were performed using a DNS solver, 
which used the fifth order WENO
finite difference scheme \citep{JS1996} to treat the convection terms of 
the governing equations and the sixth order standard central difference 
scheme to discretise the nonlinear diffusion terms. 
The time integration is the third order strong stability-preserving
Runge-Kutta method \citep{GST01}.
The advantage of the WENO finite difference method is the capability to 
achieve arbitrarily high order accuracy in smooth regions while capturing 
sharp discontinuity.

\subsection{Setup of the problem}
\label{sec:setup}
We consider an unburned gas mixture under uniform initial conditions except
for the aforementioned linear temperature gradient.
The initial conditions at $t=0$ are constant pressure and zero velocity
of the unburned mixture.
On the left boundary at $x=0$, we assume a reflecting wall,
where $U_x(x=0,t)=0$ and the initial temperature, $T(x=0)=T^{*}$ exceeds
the ignition threshold value.
Thus, the initial conditions are as follows:
\begin{subequations}
\label{eq:init}
\begin{align}
T(x,0) & =
\left\{
\begin{array}{lcc}
T^{*} - \left(T^{*} - T_0\right)\,x/L, & \quad & 0\leq x \leq L, \\[0.2em]
T_0                                    & \quad & x>L,            \\
\end{array}
\right. \\
p(x,0)&=p_0, \\
\UU(x,0)&=\bm{0}.
\end{align}
\end{subequations}

According to the Zeldovich gradient mechanism, the reactions begin
primarily at the temperature maximum, $T^{*}$, and then propagate along the
temperature gradient due to spontaneous auto-ignition of the mixture.
The velocity of the spontaneous reaction wave,
\begin{equation}\label{Usp}
U_{\rm sp}=\frac{{\rm d} x}{{\rm d}\tau_{\rm ind}}
=\left(\frac{{\rm d}\tau_{\rm ind}}{{\rm d}T}\right)^{-1}
\left(\frac{{\rm d}T}{{\rm d}x}\right)^{-1}
\end{equation}
depends on $\dd{\tau_{\rm ind}}/\dd T$ and the steepness of the
temperature gradient.
It could be larger than that of the pressure wave, if the temperature
gradient is sufficiently shallow.
Then, the coupling between the spontaneous reaction wave with the shock
wave, along with the coherent energy release in the reaction, may cause
shock wave amplification and the transition into a detonation wave.
Since we only consider the process of detonation initiation, the
parameters in equation~(\ref{eq:init}) are chosen as follows:
\begin{equation}
\begin{array}{ccccccc}
T^*=1500\K ,& & T_0=300\K,&  & L=8\,\cm,&&p_0=1\,\bbar.
\end{array}
\end{equation}
This set of parameters was also used by \cite{liberman2012regimes}
to produce a steady detonation wave in a stoichiometric hydrogen--oxygen 
mixture.

\section{Results from the {\sc Pencil Code}}

\subsection{General remarks regarding the transition to detonation (TD)}

In the absence of shock viscosity, or when the shock viscosity is too small,
small-scale oscillations on the grid scale (wiggles) occur.
Such a solution cannot be numerically reliable and must be discarded.
When we add shock viscosity, the wiggles become weaker.
However, when the shock viscosity is too large, TD is no longer possible.
Thus, to pose a meaningful convergence test, we decided to fix the value
of $C_{\rm shock}$ to a relatively small value of $0.8$ and then increase
the resolution.
This means that the shock viscosity continuously decreases with increasing
resolution until it becomes negligible.

\begin{table}[t!]\caption{
Summary of the fit parameters at $t=42\us$;
$x_0$ is in cm, $p_0$ and $p_1$ are in bar, $p_1'$ is in $\bbar\um^{-1}$,
$L_1$ and $L_2$ are in $\um$, and $\delta t_{\min}$ in $\ps$.
Runs~(a)--(d) have $C_{\rm shock}=0.8$ and run~(e) has $C_{\rm shock}=0.2$.
}\vspace{12pt}\centerline{\begin{tabular}{cccccccrrc}
& $\delta x$ & $x_0$ & $p_1$ & $p_1'$ & $L_1$ & $L_2$ & $N_x\quad$  & $N_t\quad$ & $\delta t_{\min}$ \\
\hline
(a) & 1.993 & 9.37498 & $35.00$ & $0.2200$ & 2.56  & 0.36   &  50,176 &   392,000 & $42$ \\
(b) & 0.997 & 9.44825 & $31.20$ & $0.0546$ & 1.21  & 0.30   & 100,352 & 1,266,600 & $24$ \\
(c) & 0.498 & 9.44390 & $33.21$ & $0.0535$ & 0.442 & 0.0587 & 200,704 & 2,826,300 & $12$ \\
(d) & 0.199 & 9.27530 & $35.78$ & $0.1128$ & 0.1145& 0.0157 & 501,760 &14,603,000 & $2.5$ \\
(e) & 0.199 & 9.46444 & $28.70$ & $0.0300$ & 0.4069& 0.1719 & 501,760 &14,255,800 & $2.5$ \\
\label{Tsummary}\end{tabular}}\end{table}

\begin{figure*}[t!]\begin{center}
\includegraphics[width=\textwidth]{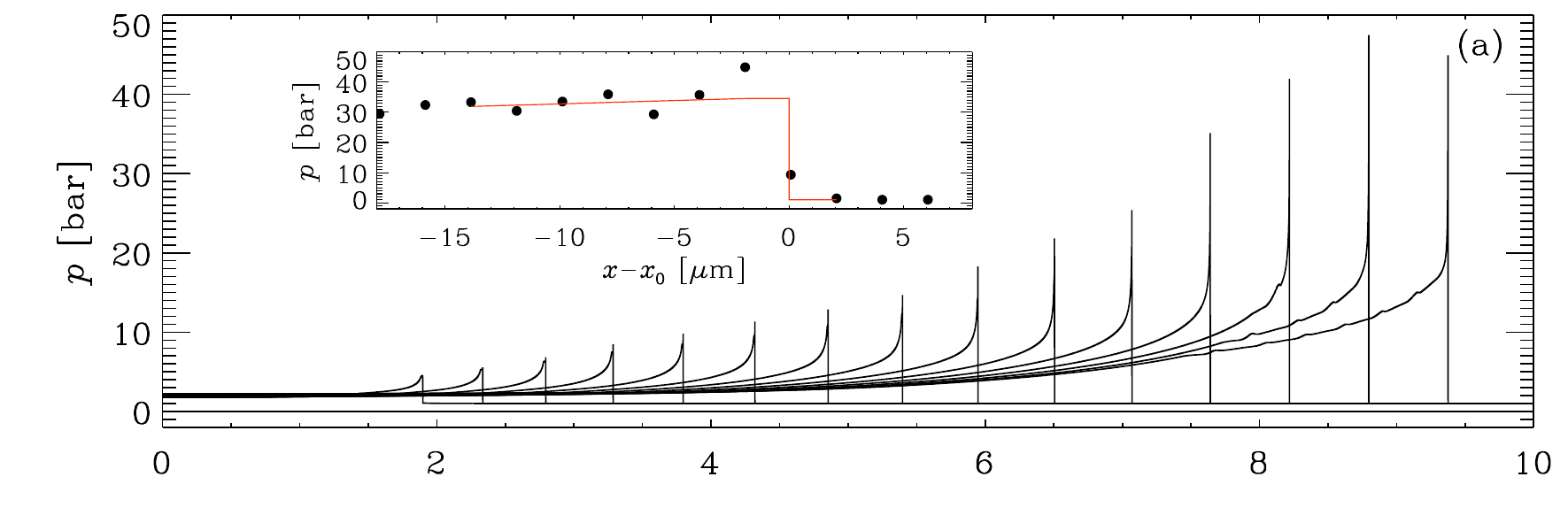}
\includegraphics[width=\textwidth]{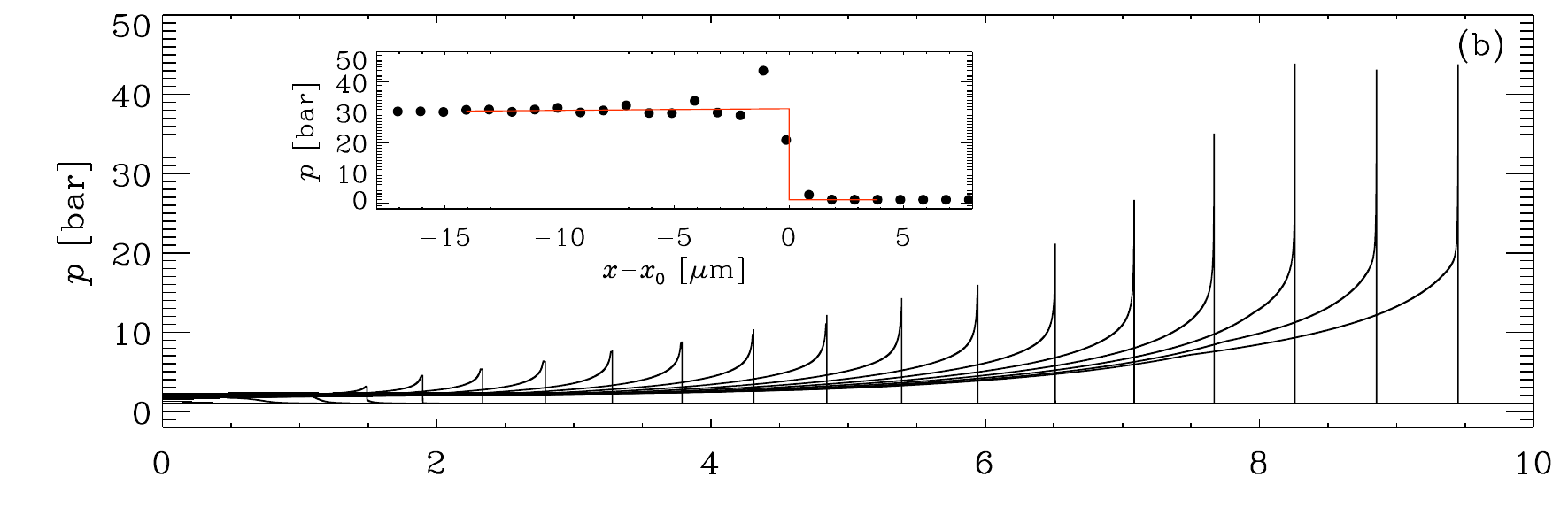}
\includegraphics[width=\textwidth]{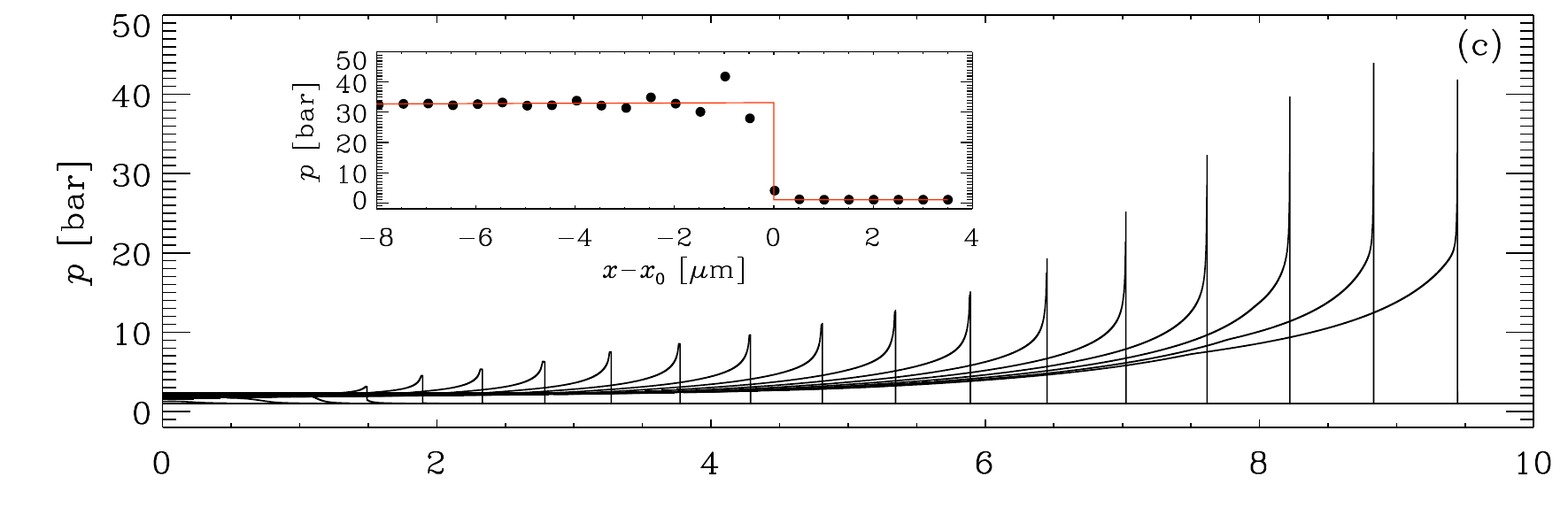}
\includegraphics[width=\textwidth]{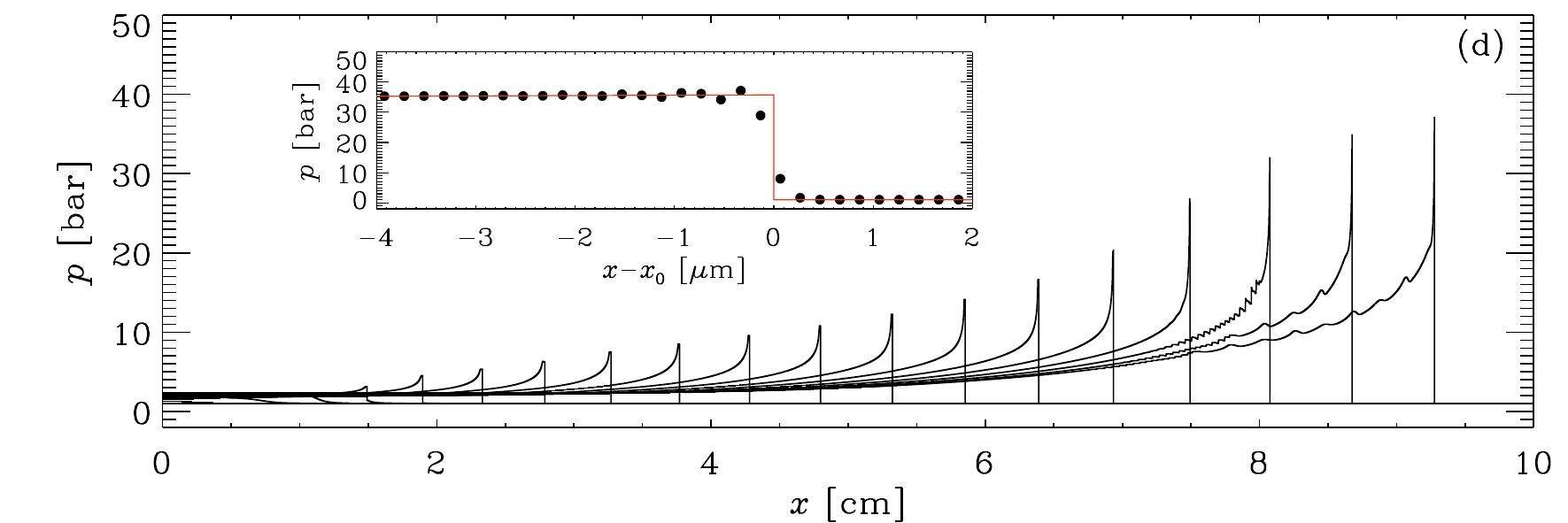}
\end{center}\caption[]{
Pressure profiles for
(a) $\delta x=2\um$, (b) $1\um$, (c) $0.5\um$, and (d) $0.2\um$ in
regular time intervals from $t=2\us$ to $42\us$.
The insets show the pressure peak at the last time, indicated by filled
symbols, where the red line shows the fit in the proximity of the
pressure peak.
Note that the $x$ range varies (colour online).
}\label{ppvar_range_Hotspot_shockdiff_nx501760b}
\end{figure*}

\begin{figure*}[t!]\begin{center}
\includegraphics[width=\textwidth]{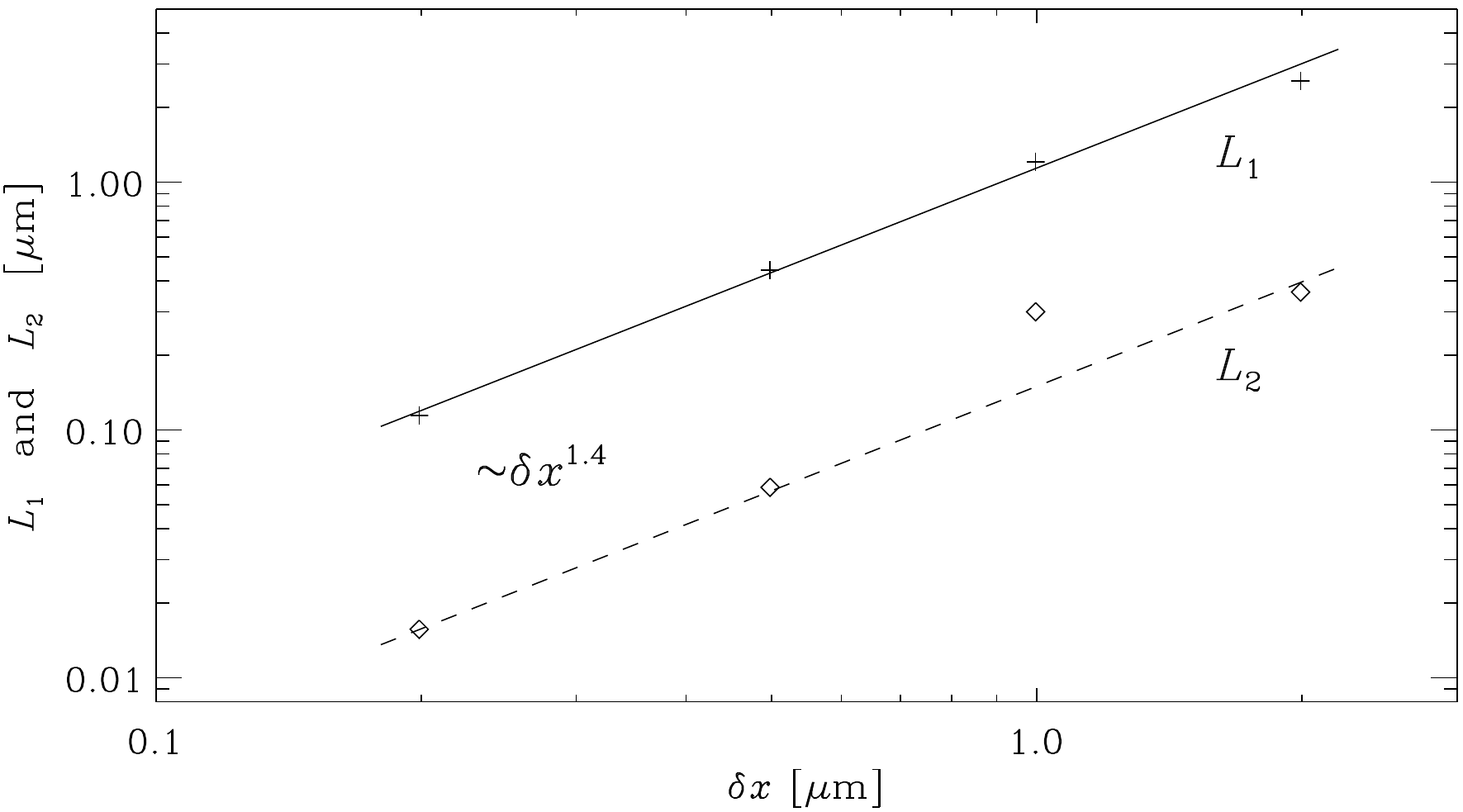}
\end{center}\caption[]{
Convergence of $L_1$ and $L_2$ with $\delta x$.
}\label{pconvergence}\end{figure*}

\subsection{The pressure wave at increasing resolution}

With each doubling of the number of mesh points, the total shock viscosity
integrated over the width of the shock decreases by a factor of four.
In addition, there is the time-dependent molecular viscosity profile
which is independent of the mesh resolution.
Thus, we expect that in the limit of infinite resolution, 
which yields a vanishing shock viscosity, the wiggles
of the tip of the pressure profile should disappear.
To test this assertion, and to study the corresponding convergence
property of the code, we perform four simulations with mesh resolutions
between $\delta x=2\um$ and $0.2\um$ using $C_{\rm shock}=0.8$;
see \Tab{Tsummary}.
The corresponding pressure profiles are shown in
\Fig{ppvar_range_Hotspot_shockdiff_nx501760b} for the four cases (a)--(d).
The insets in each panel show the corresponding pressure profile
in the proximity of the peak.

It is evident that the wiggles decrease as we increase the resolution.
In addition, the pressure profiles change slightly with resolution.
To characterise these changes, we determine a fit to the pressure
peak of the form
\EQ
p_{\rm fit}(x, t_\ast)=p_0+\left[p_1+(x-x_0)\,p_1'\right]\,
\varTheta(x_0-x),
\label{pfit}
\EN
where $\varTheta(x)$ is the Heaviside step function ($=1$ for $x>0$ and zero
otherwise), $x_0$ is the position of the peak at the last time $t_\ast$,
$p_0$ is the atmospheric background pressure ahead of the peak,
$p_1$ is the pressure increase relative to $p_0$ just behind the peak,
and $p_1'$ is the slope of the pressure profile to the left of the peak,
i.e., in the wake of the detonation wave.
In all cases, the pressure ahead of the peak is
$p_0=1.013\times10^6\dyn\cm^{-2}$, which does not need to be fitted.
The remaining three parameters are given in \Tab{Tsummary},
where the pressure is given in bar ($1\bbar=10^6\dyn\cm^{-2}$).

Note that between the runs with $\delta x=1\um$ and $0.5\um$,
the front speeds (or front positions $x_0$) agree within 0.05\%, but for the run with
$\delta x=0.2\um$, the front speed has decreased by nearly 2\%.
The reason for this apparent loss of accuracy is not fully identified,
although it is clear that smaller values of $C_{\rm shock}$ lead again
to larger front speeds; see run~(e) in \Tab{Tsummary}.
It is therefore possible that at this high resolution, the value
$C_{\rm shock}=0.8$ is already too large and that a smaller value,
for example around $0.6$, could be more reasonable.
We should also point out that we have used a relatively optimistic choice
of the viscous time step (we chose $\delta t\,\nu_{\max}/\delta x^2=0.4$
instead of the more conservative value of 0.25 that is recommended in
the manual to the {\sc Pencil Code}).
However, comparisons with the smaller value did not indicate any differences
in the front speed.
The fact that the viscous time step enters in this highest resolution
run, but not in the others, is related to the extremely small mesh size
in this case.
This makes the time step constraint from the relatively large molecular
viscosity near $x=0$ very severe.
Note also that in this run, waves appear in the wake of the pressure
field behind the peak after $t=36\us$.
These also seem to be spurious and are not found when $C_{\rm shock}$
is smaller.

Next, to characterise the convergence, we use the $L_1$ and $L_2$ norms
defined here as follows:
\EQ
L_1=\int_{x_1}^{x_2}\left|p(x,t_{\ast})-p_{\rm fit}(x,t_{\ast})\right|\,\dd x/(p_0+p_1),
\EN
\EQ
L_2=\int_{x_1}^{x_2}\left|p(x,t_{\ast})-p_{\rm fit}(x,t_{\ast})\right|^2\,\dd x/(p_0+p_1)^2.
\EN
Both have the dimension of a length.
These values are also given in \Tab{Tsummary}.
Figure~2 shows that $L_1$ and $L_2$ decrease with resolution like $\delta x^{1.4}$.

\begin{figure*}[t!]\begin{center}
\includegraphics[width=\textwidth]{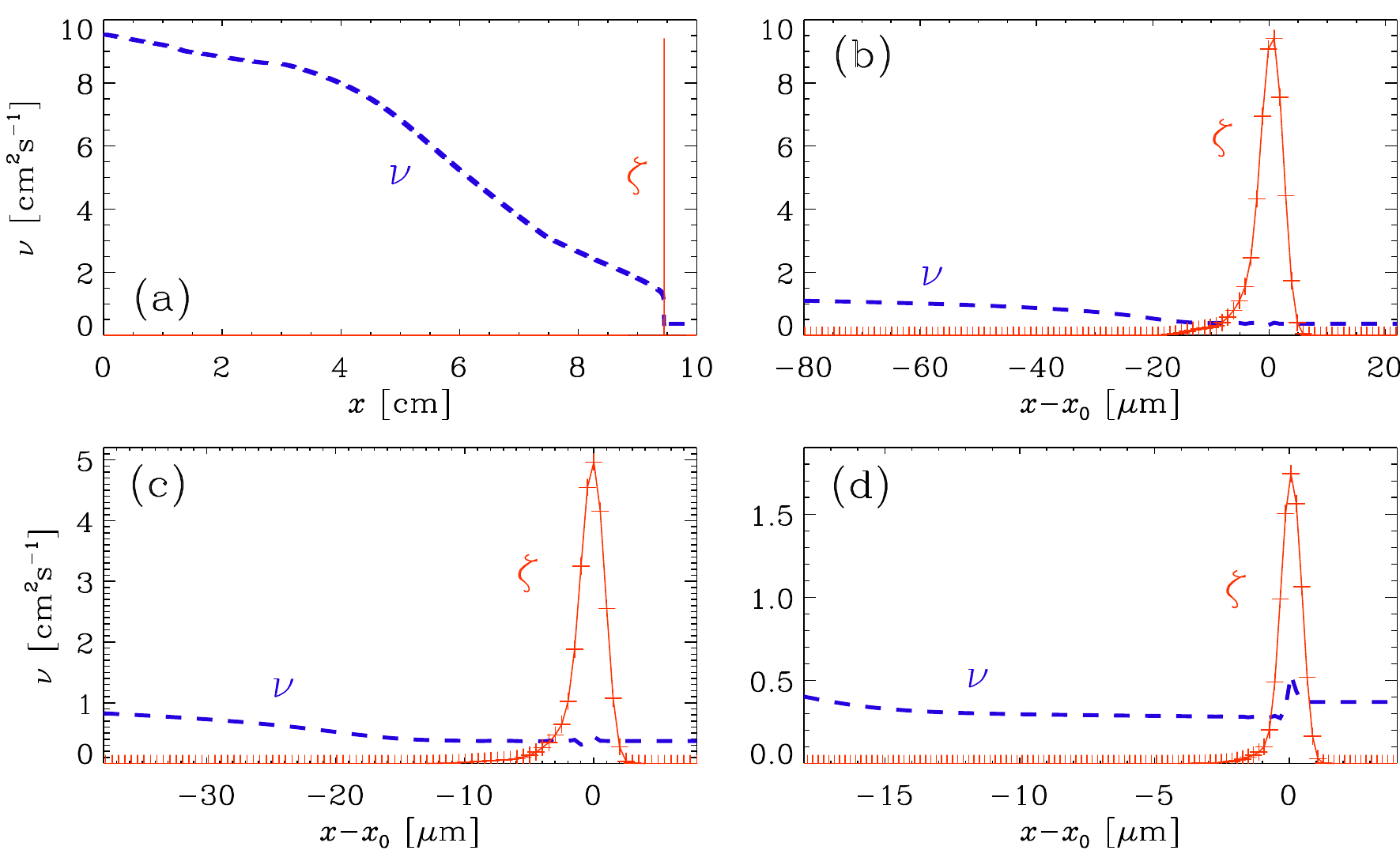}
\end{center}\caption[]{
Comparison of the profiles of viscosity ($\nu$, blue dashed lines) and
shock viscosity ($\zeta$, red lines with mesh points being marked with
plus signs) for (a) $\delta x=1\um$ showing the full $x$ range from $0$
to $10\cm$.  (b)--(d) show only the close proximity of the shock at $x_0$
for (b) $\delta x=1\um$, (c) $\delta x=0.5\um$, and (d) $\delta x=0.2\um$,
at $t=42\us$ (colour online).
}\label{ppvar_comp_nu}\end{figure*}

In \Fig{ppvar_comp_nu} we compare the molecular viscosity profile 
at the last time with the corresponding shock viscosity 
for the three highest resolutions shown in
figures~\ref{ppvar_range_Hotspot_shockdiff_nx501760b}(b)--(d).
The overall profile of the molecular viscosity is the same in all three
cases and varies significantly from $\sim10\cm^2\s^{-1}$ at $x=0$ to
$\sim0.3\cm^2\s^{-1}$ at and ahead of the shock.
However, the peak of the shock viscosity
decreases from $\sim10\cm^2\s^{-1}$ in
figure~\ref{ppvar_range_Hotspot_shockdiff_nx501760b}(b)
by about a factor of five to $\sim1.8\cm^2\s^{-1}$ in
figure~\ref{ppvar_range_Hotspot_shockdiff_nx501760b}(d).
In addition, the width of the shock viscosity profile also decreases by
about a factor of five, so the integrated effect of the shock viscosity
diminishes by a factor of about 25, as expected from \Eq{zeta_shock}.
Note that for the highest resolution, the shock viscosity makes up
a small contribution compared to the molecular viscosity.
We emphasise that the maximum of the molecular viscosity
($\sim10\cm^2\s^{-1}$ at $x=0$) is much larger than the maximum of the
shock viscosity ($\sim1.8\cm^2\s^{-1}$ at $x=9.4\cm$, although at this
point the molecular value is only $\sim0.3\cm^2\s^{-1}$); compare
figures~\ref{ppvar_comp_nu}(a) and \ref{ppvar_comp_nu}(d).
This means that at late times, when $\nu$ has become large far in the
wake of the shock, an enormous amount of time is spent because the
viscous time step is then so short.
This is also evident from \Tab{Tsummary}, where we see that the total
number of time steps has increased by a factor of over five as the
resolution was increased by only a factor of $2.5$.

\begin{figure*}[t!]\begin{center}
\includegraphics[width=\textwidth]{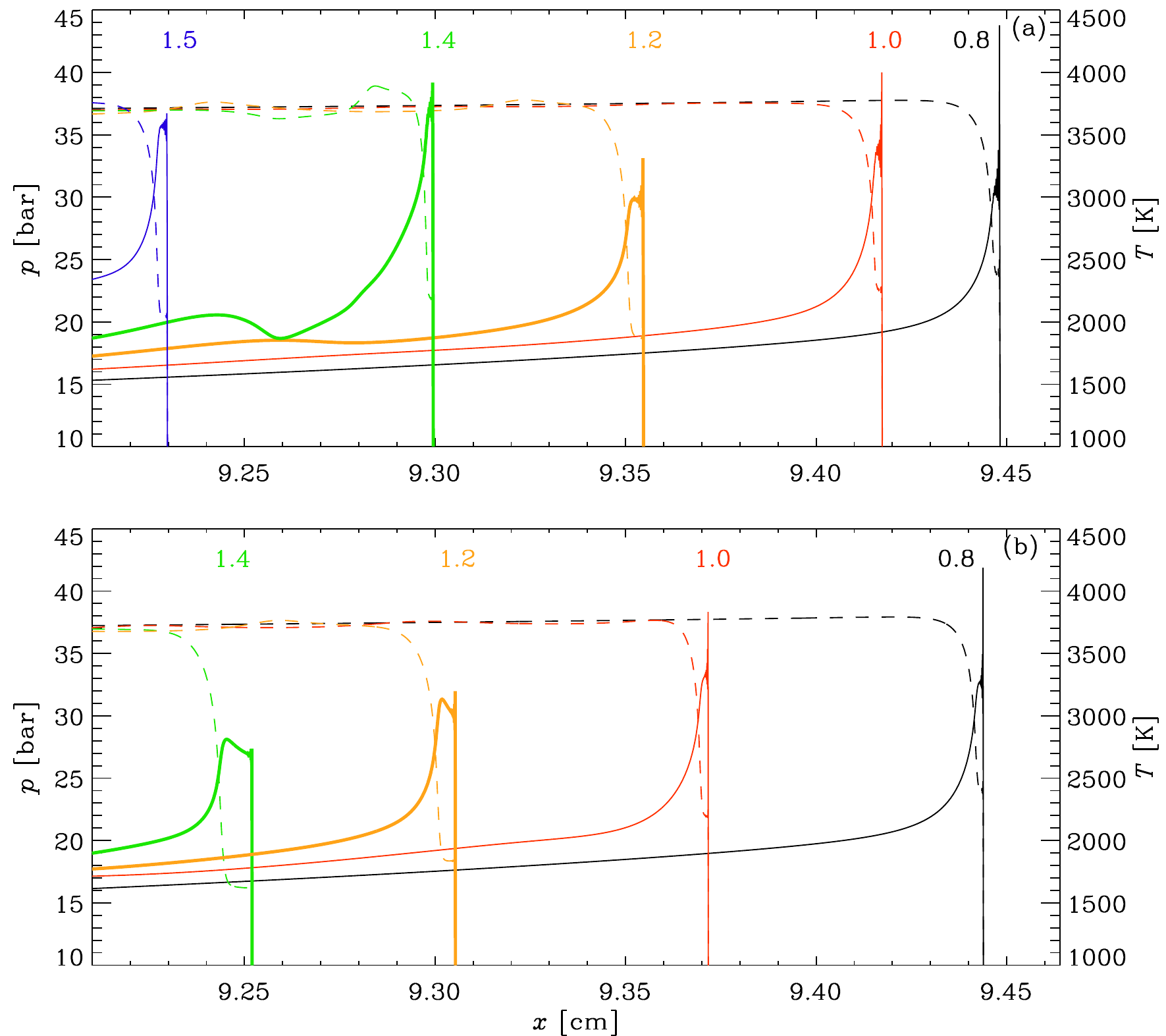}
\end{center}\caption[]{
Comparison of pressure and temperature profiles for $C_{\rm shock}=0.8$ (black),
$1.0$ (red), $1.2$ (orange), $1.4$ (green), and $1.5$ (blue),
for $\delta x=1\um$ (top) and $0.5\um$ (bottom).
Note that for $C_{\rm shock}=1.5$ and $\delta x=0.5\um$, no TD
develops (colour online).
}\label{ppvar_comp_2panels}\end{figure*}

\begin{figure*}[t!]\begin{center}
\includegraphics[width=\textwidth]{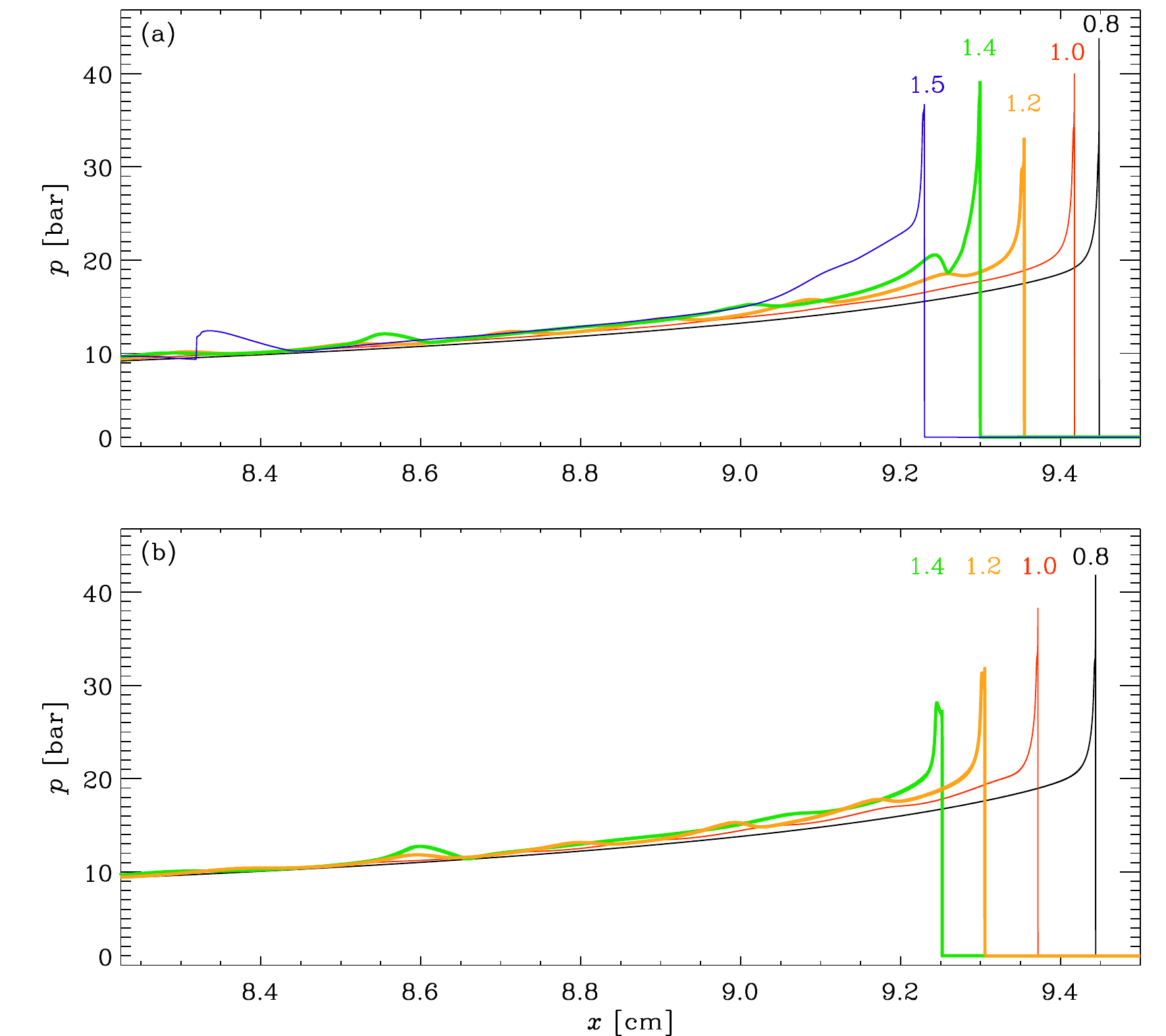}
\end{center}\caption[]{
Similar to \Fig{ppvar_comp_2panels}, but showing also the
wake of the pressure front.
Note the waves in the simulations with $C_{\rm shock}>0.8$;
see the blue, green, orange, and red lines (colour online).
}\label{ppvar_comp_2panels_wake}\end{figure*}

\subsection{Dependence on $C_{\rm shock}$}

Next, we investigate the dependence of our solutions on the value of
$C_{\rm shock}$.
In \Fig{ppvar_comp_2panels} we show pressure profiles for different
values of $C_{\rm shock}$ at resolutions of $\delta x=1\um$ and $0.5\um$.
For $C_{\rm shock}\leq1.4$, the pressure profiles still show wiggles at
the position of the pressure maximum, but the wiggles are smaller and
more localised at the higher resolution of $0.5\um$.
For $C_{\rm shock}=1.4$, however, the wiggles are nearly completely
negligible at the resolutions of $0.5\um$, but the pressure profile has
also changed in that case and has now a short flank with a negative
slope just behind the shock, that is, to its left.
For $C_{\rm shock}=1.5$, TD is only found in the case with
$\delta x=1\um$, but not with $\delta x=0.5\um$.

In \Fig{ppvar_comp_2panels_wake}, we show a larger portion of the
wake behind the pressure front, where we see the occurrence of
another type of long-wavelength oscillation, when $C_{\rm shock}$
is larger than $0.8$.
Those waves are similar for both the higher and lower resolution runs,
but could also be a feature of having under-resolved the solution
at earlier times that are not shown here.

\subsection{Speeds of pressure and reaction fronts}

\begin{figure*}[t!]\begin{center}
\includegraphics[width=\textwidth]{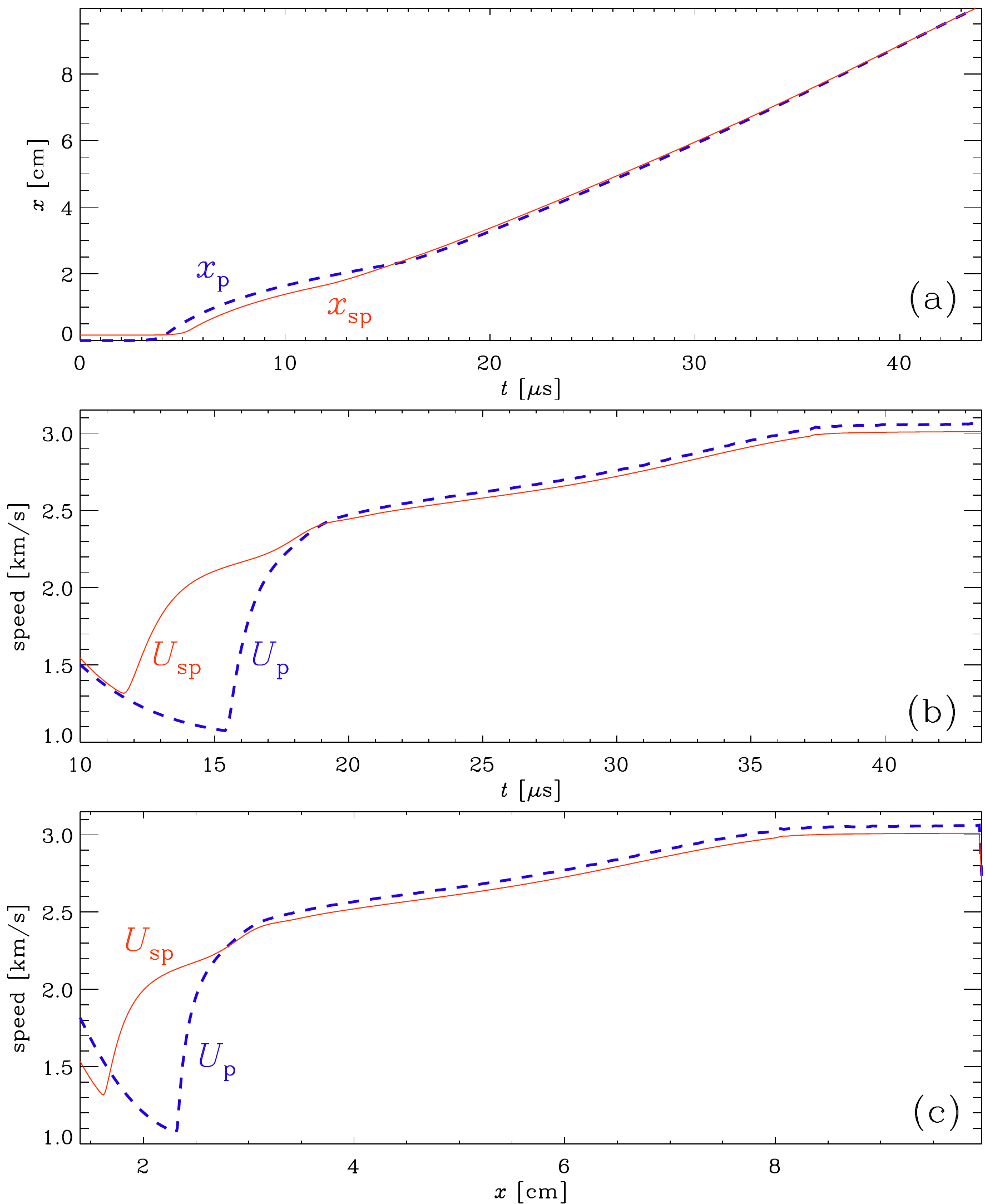}
\end{center}\caption[]{
Front positions and speeds for $\delta x=0.5\um$ and $C_{\rm shock}=0.8$.
(a) $x_{\rm sp}(t)$ (red solid line) and $x_{\rm p}(t)$ (blue dashed line),
(b) $U_{\rm sp}(t)$ (red solid line) and $U_{\rm p}(t)$ (blue dashed line), and
(c) $U_{\rm sp}(x)$ (red solid line) and $U_{\rm p}(x)$ (blue dashed line)
(colour online).
}\label{pspeed_vs_x_Hotspot_shockdiff_nx200704a_rep}\end{figure*}

Finally, we show in \Fig{pspeed_vs_x_Hotspot_shockdiff_nx200704a_rep}
the time dependence of the positions and speeds of the pressure and
spontaneous reaction fronts.
In practice, the speeds $U_i$ (with $i={\rm p}$ for pressure and $i={\rm
sp}$ for spontaneous reaction wave) are computed by time differentiation
of the position $x_i$, which is obtained from the volume where the
pressure or the reactant are above a certain threshold.
Specifically, we compute
\EQ
U_{\rm p}={\dd x_{\rm p}\over\dd t}
=-{\dd\over\dd t}\int_0^{x_{\max}}
\max(p_{\rm crit}-p,\,0)/(p_{\rm crit}-p_0)\,\dd x,
\EN
where we have used $p_{\rm crit}=1.020\bbar$ as threshold pressure and
$p_0=1.013\bbar$ is still the same background pressure as in \Eq{pfit}.
The spontaneous reaction speed is based on the amount of water produced, i.e.,
\EQ
U_{\rm sp}={\dd x_{\rm sp}\over\dd t}
=-{\dd\over\dd t}\int_0^{x_{\max}}
\max(1-Y_k/Y_{k0},\,0)\,\dd x,
\EN
where $k={\rm H}_2{\rm O}$ and $Y_{k0}=0.3$ is half the value
of $Y_k\approx0.6$ after ${\rm H}_2$ has reacted with ${\rm O}_2$.
The final values of the two speeds are
$U_{\rm p}=3.06\km\s^{-1}$ and $U_{\rm sp}=3.01\km\s^{-1}$.
These values are close to the empirically determined value of $3.0\km\s^{-1}$,
which is only known to within about 1\% accuracy and therefore compatible
with our results.

According to \Eq{Usp}, the velocity of the spontaneous
reaction wave decreases in the beginning, since
$U_{\rm sp} \propto (\dd \tau_{\rm ind}/\dd T)^{-1}$.
It reaches a minimum somewhere near the crossover temperature
$T_{\rm cr} \approx 1000\K$ (for the present mixture of ${\rm H}_2$ and
${\rm O}_2$) for the steepest gradient capable of initiating detonation
which corresponds to the transition from the endothermal to the exothermal
stage of the reaction.
In our case, the gradient is rather shallower, so the minimum of the
velocity is reached earlier.

After reaching the minimum velocity, the speed of the spontaneous reaction
wave increases due to energy release in the reaction.
To accomplish coupling between the spontaneous reaction wave and the pressure
wave, it is necessary (but not sufficient) that $U_{\rm sp}>U_{\rm p}$
after the point where $U_{\rm sp}$ is minimum, which
is the case during the interval $12\us\leq t\leq19\us$.
For $t>19\us$, the coupling between the reaction wave and the shock wave
is developing until detonation is reached at $t\approx38\us$.
Note also that $U_{\rm sp}$ is now slightly less than $U_{\rm p}$,
but this is natural because the reaction happens always slightly behind
the leading shock.
In fact, at late times, hardly any difference
between $x_{\rm p}$ and $x_{\rm sp}$ can be seen; see
\Fig{pspeed_vs_x_Hotspot_shockdiff_nx200704a_rep}(a).
This is compatible with the experimental value of the detonation;
see \cite{Kuz_etal05}.

\section{Comparison with the WENO code}

\begin{figure*}[t!]\begin{center}
\includegraphics[width=0.49\textwidth]{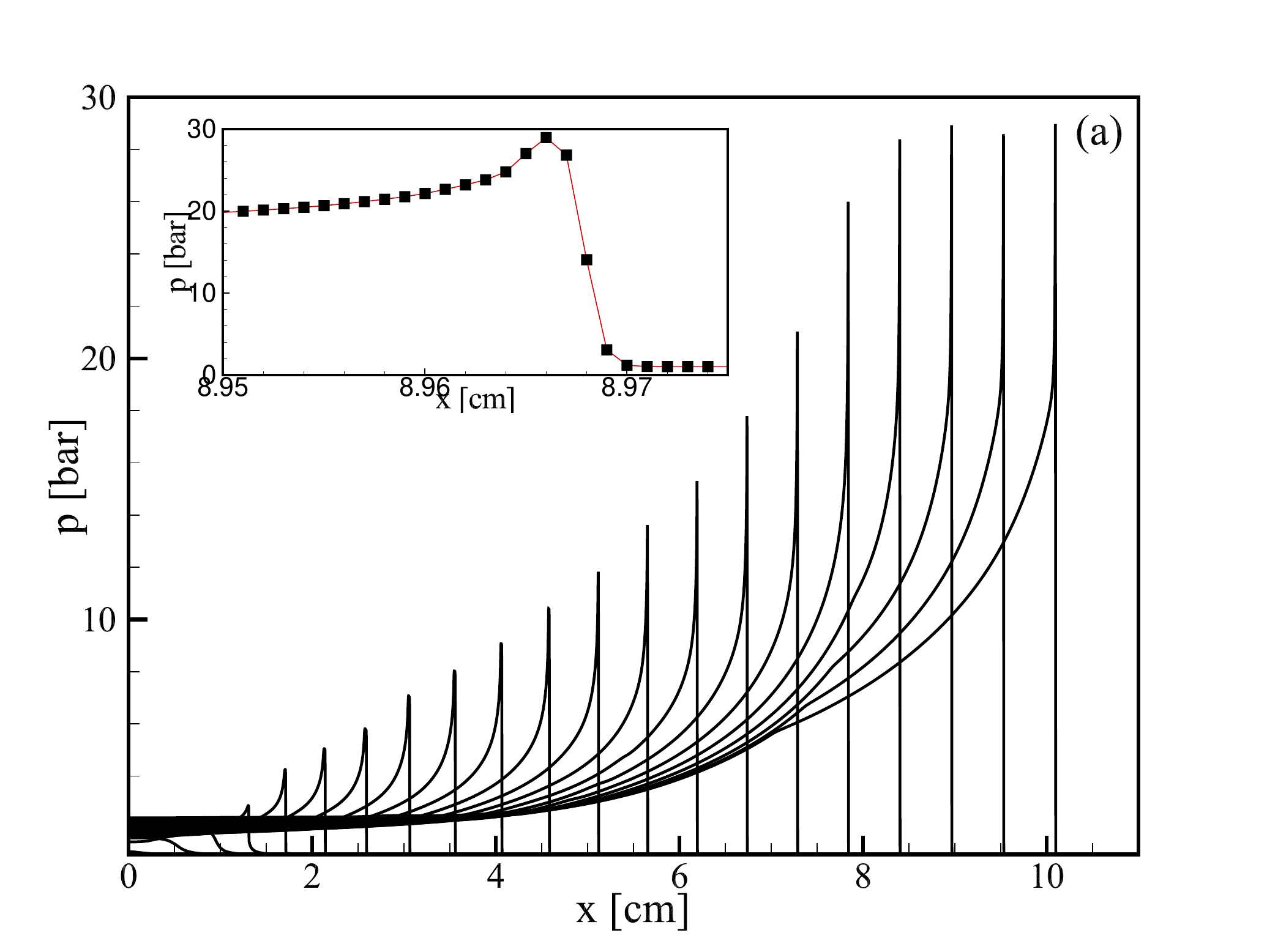}
\includegraphics[width=0.49\textwidth]{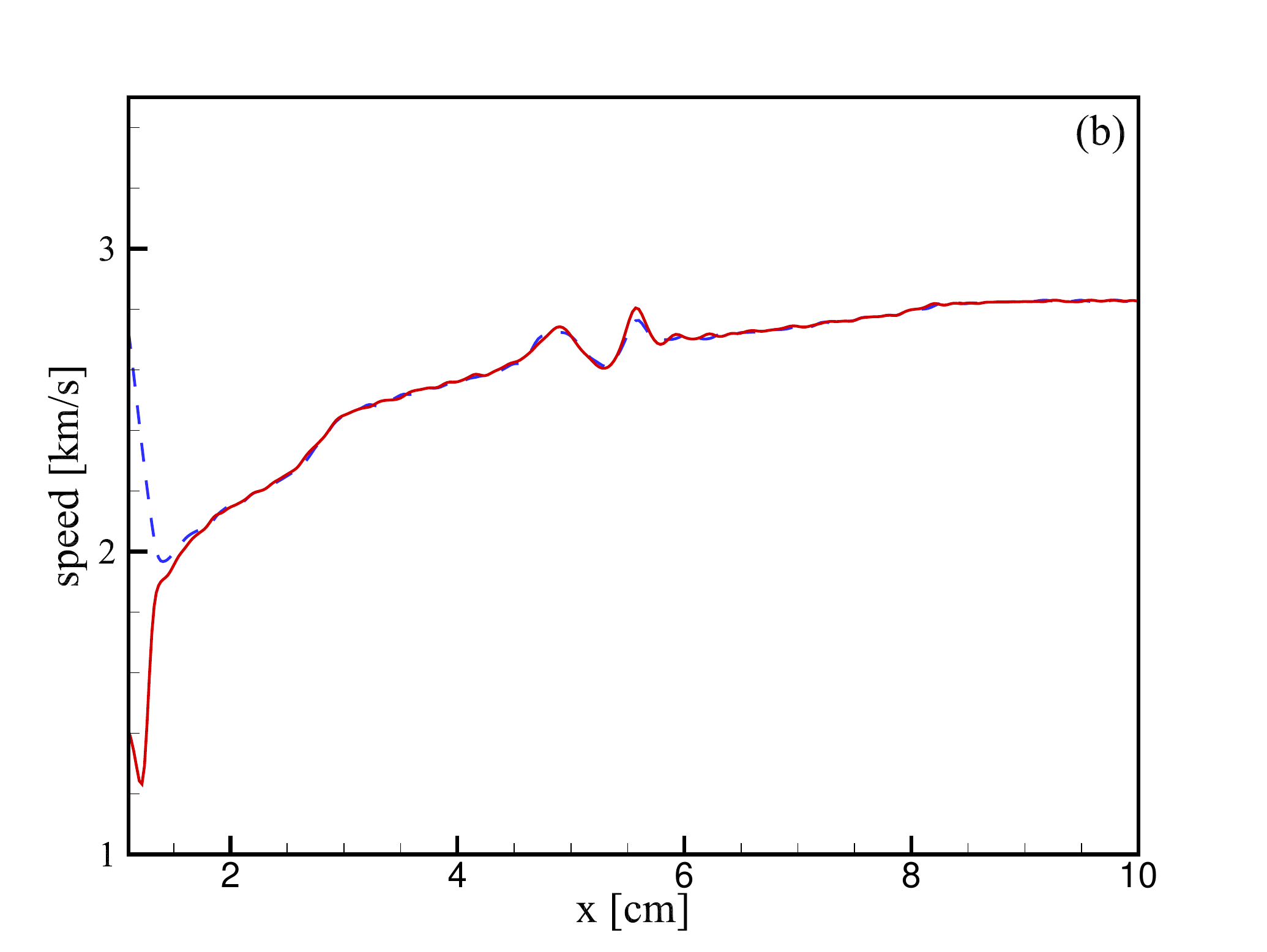}
\end{center}\caption[]{(a) pressure profiles calculated with the WENO code
at resolution $\delta x=10\um$, in regular time intervals from $0\us$ to 
$46\us$. The inset shows the vicinity of the pressure peak at $42\us$. (b) 
corresponding spontaneous wave velocity (red solid line) and pressure wave 
velocity; see the blue dashed line (colour online).
}\label{WENO_dx_10um}\end{figure*}

\begin{figure*}[t!]\begin{center}
\includegraphics[width=0.49\textwidth]{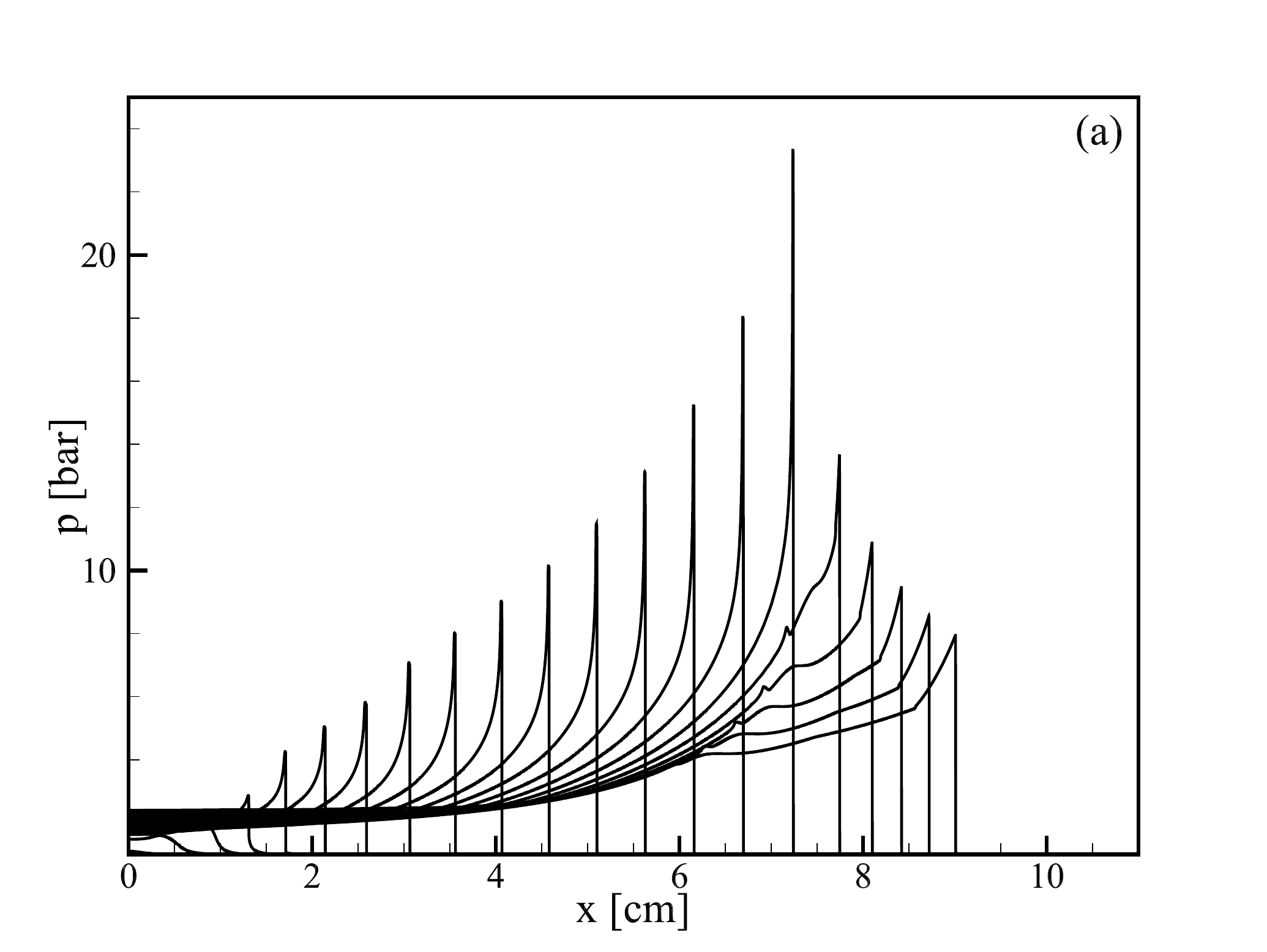}
\includegraphics[width=0.49\textwidth]{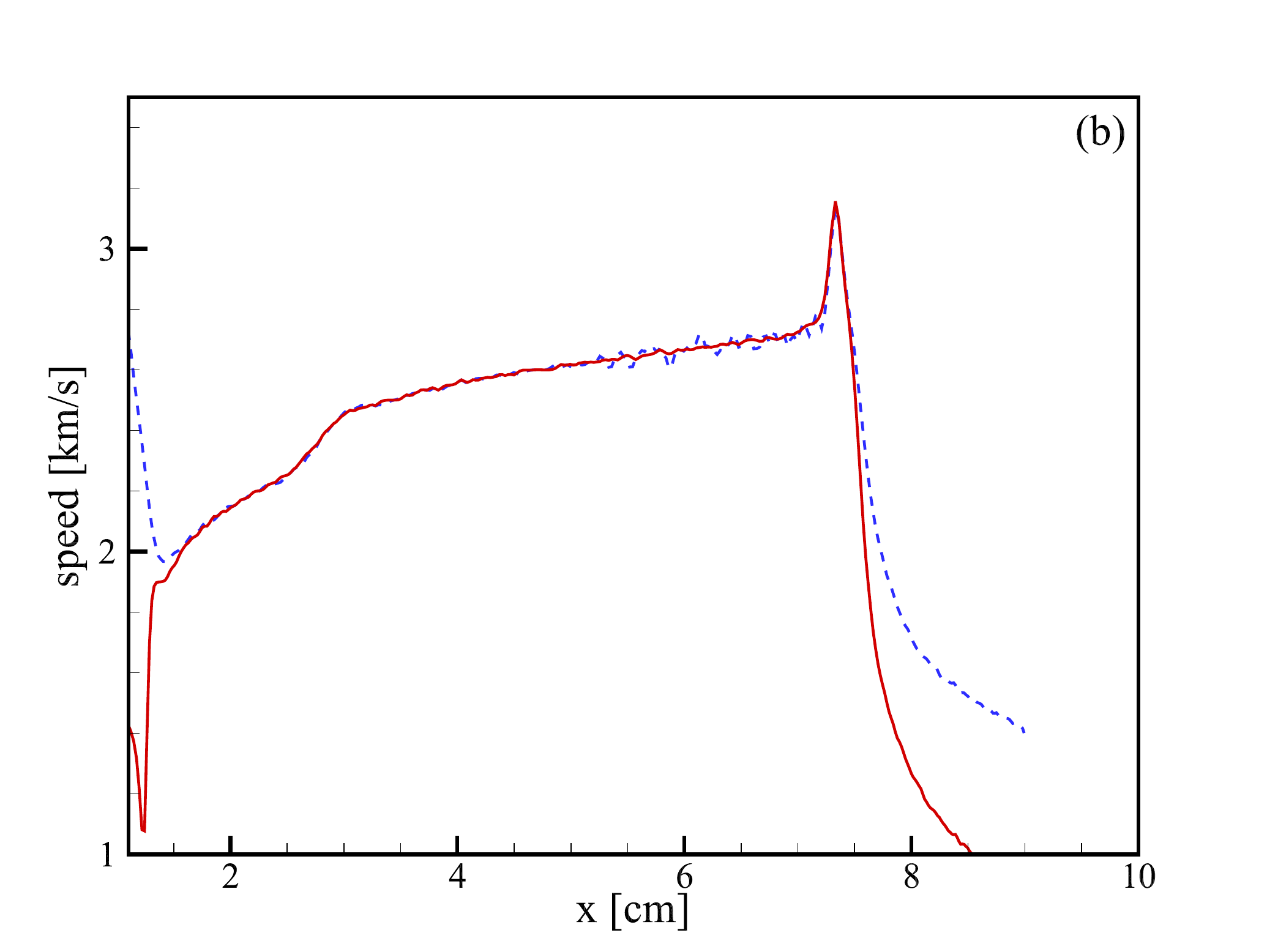}
\end{center}\caption[]{
Similar to \Fig{WENO_dx_10um}, but for $\delta x=5\um$ and without inset
(colour online).
}\label{WENO_dx_5um}\end{figure*}

There is extensive literature devoted to the simulation of hydrodynamic 
problems with shock and detonation waves using a shock-capturing approach 
\citep{HS2019,DFZLL2019,FZMZ2019,DXTX2019,DX2019,CAYMZ2018,Zhao2018}. 
In this section we consider solutions to the detonation problem (see 
\Sec{sec:Shocks} for details) using the WENO code, which is widely used to 
simulate various combustion and detonation problems. 
Compared to the results obtained with the {\sc Pencil Code}, there are no wiggles in the 
pressure profiles without the addition of a shock viscosity due to the usage of 
the WENO scheme. 
Figure~\ref{WENO_dx_10um}(a) shows the evolution of pressure profiles 
during the formation of a steady detonation after the coupling of the 
spontaneous reaction and shock waves has been obtained in the simulations with the WENO 
code at resolution $\delta x=10\um$. 
The corresponding spontaneous reaction wave velocity and pressure wave 
velocity are presented in figure~\ref{WENO_dx_10um}(b). 
Small oscillations of the velocities of the reaction and pressure waves 
indicate the coupling of shock and spontaneous reaction
waves in the beginning of the development toward detonation.
However, the simulations shown in figure~\ref{WENO_dx_5um} at a higher 
resolution with $\delta x=5\um$ show that the developing detonation quenches 
before it leaves the temperature gradient. 
The previously successfully coupled reaction and shock waves are decoupled 
at around $7.4\cm$. 
It is worth noting that the quenching of detonation in this case is in no 
way due to the gradient being too steep. 
Simulations with a resolution of $\delta x=5\um$ for a much shallower gradient 
($L=18\cm$) also show that the initially coupled reaction and shock 
waves later decouple and the initially developing detonation quenches.

\begin{figure*}[t!]\begin{center}
\includegraphics[width=0.49\textwidth]{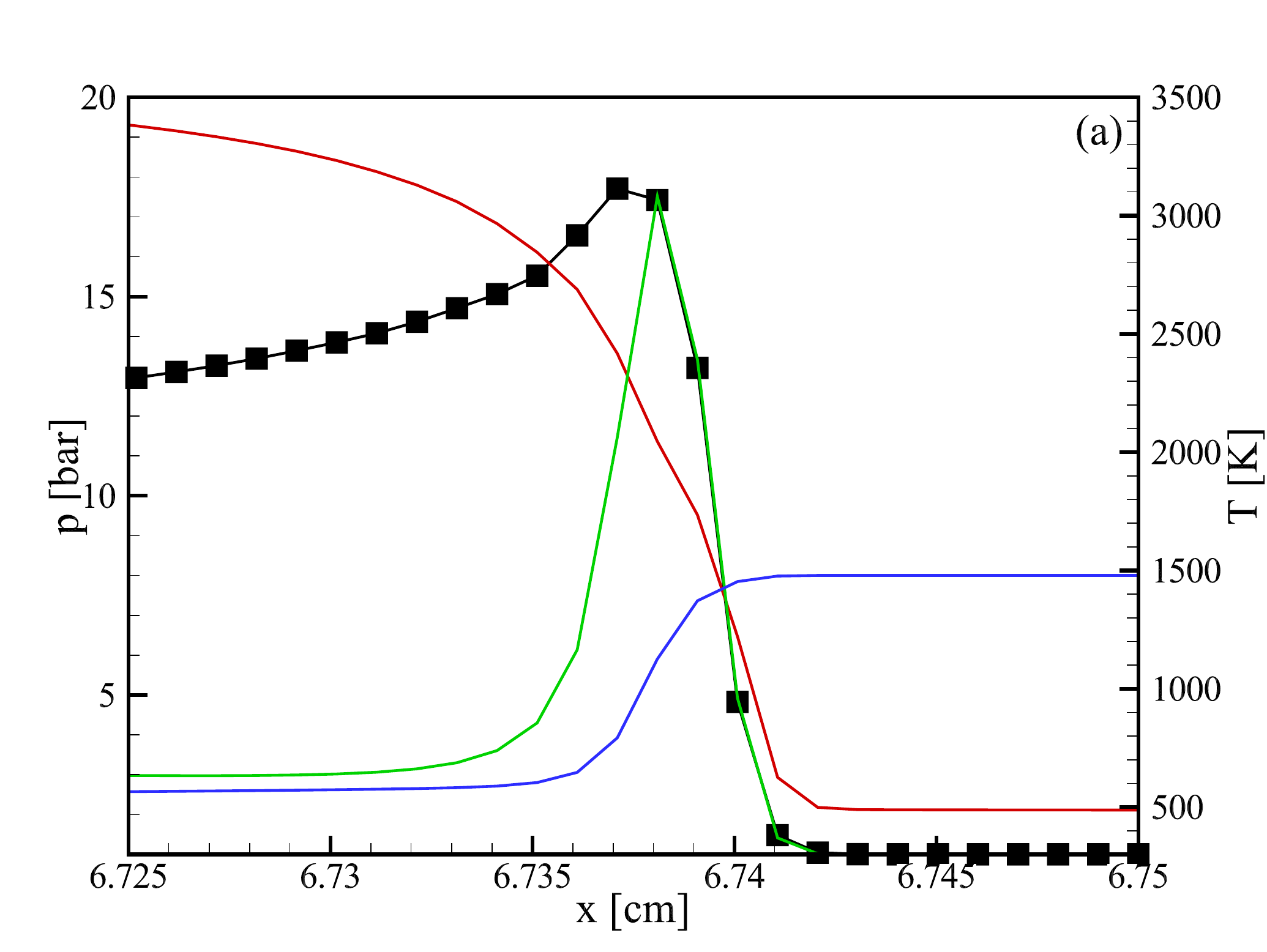}
\includegraphics[width=0.49\textwidth]{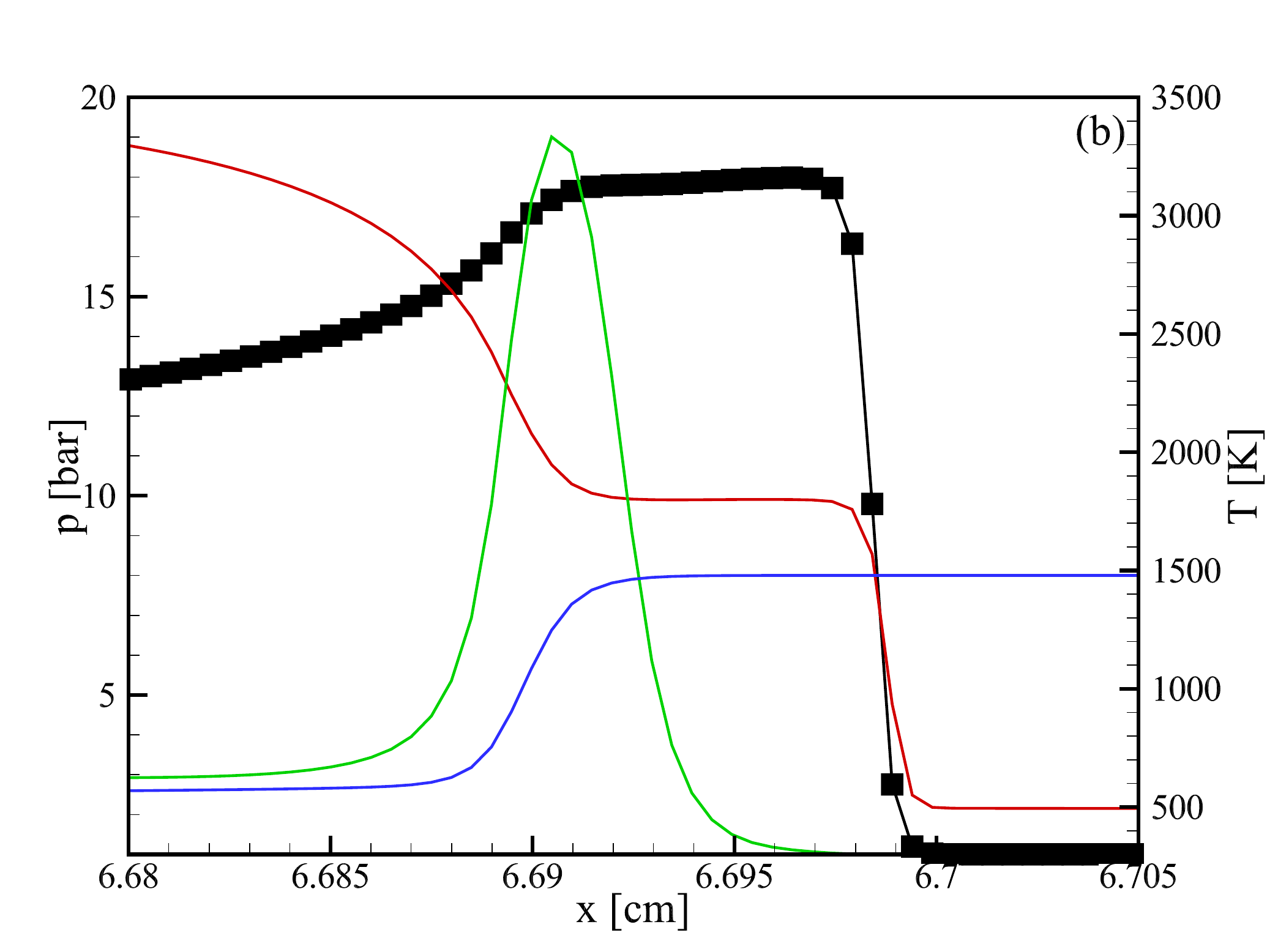}
\end{center}\caption[]{Profiles of pressure (black line), temperature (red 
line), mass fraction of ${\rm HO_2}$ (green line), and 
${\rm H_2}$ (blue line) at $t=34\us$, calculated with the WENO code 
at resolutions $\delta x=10\mu m$ (a) and $\delta x=5\mu m$ (b)
(colour online).
}\label{WENO_t_34us}\end{figure*}

Figures~\ref{WENO_t_34us}(a) and (b) show profiles of pressure, temperature,
and mass fractions of ${\rm HO_2}$ and ${\rm H_2}$ at $t = 34\us$,
calculated using the WENO scheme at resolutions $\delta x=10\um$ and 
$\delta x=5\um$. 
It is seen that, without artificial viscosity, the width of the shock is too 
small at a resolution of $\delta x=5\um$ so that the coupling of 
the reaction and shock waves becomes impossible, resulting in a quenching of 
the detonation, as shown in figure~\ref{WENO_dx_5um}. 
While the non-oscillating shock-capturing WENO scheme works quite well for 
simulations of hydrodynamic problems with shock waves, it does not work for the 
problem of detonation development, which is more "sensitive to the resolution"
compared to ordinary supersonic flows with shock waves. 
The solution obtained with the WENO scheme at a low resolution in 
figure~\ref{WENO_dx_10um} ($\delta x=10\mu m$) shows the development of a steady 
detonation, but at the higher resolution of figure~\ref{WENO_dx_5um} 
($\delta x=5\um$), the shock becomes too thin (figure~\ref{WENO_t_34us}), 
and thus could not couple with the reaction wave, so the detonation quenches. 

The physical problem in question, also known as the shock wave 
amplification by coherent energy release (SWACER) mechanism, which is a particular
case of a detonation initiated by shallow temperature (or reactivity) gradients, 
has been studied experimentally by \cite{LKY1980}. 
In this case, the shock-capturing approach of WENO does not work. 
More precisely, it works only at low resolutions, here with $\delta x=10\um$, 
when the width of the shock, obtained with WENO, is sufficiently large for 
coupling of pressure 
wave and the subsequent shock with the spontaneous reaction wave.
In simulation of the SWACER problem, shock-capturing and artificial viscosities
(numerical dissipation) must be compatible with the size of the computation 
resolution in the sense that, if the reaction wave was coupled with 
the pressure wave and later with the shock wave, it must remain coupled with 
the shock at all times until a strong shock wave is formed and then develops
into a steady detonation. 

\begin{figure*}[t!]\begin{center}
\includegraphics[width=0.49\textwidth]{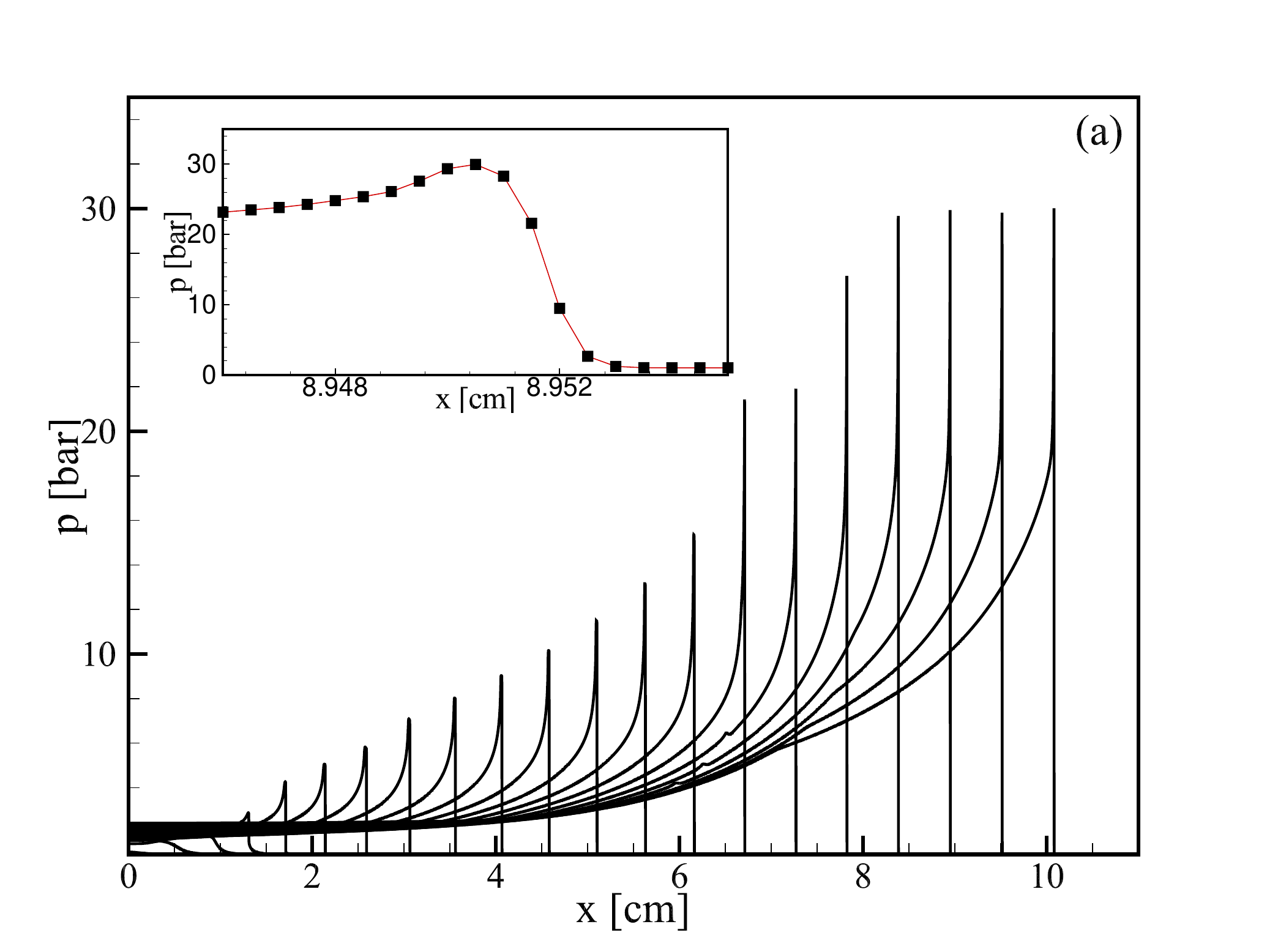}
\includegraphics[width=0.49\textwidth]{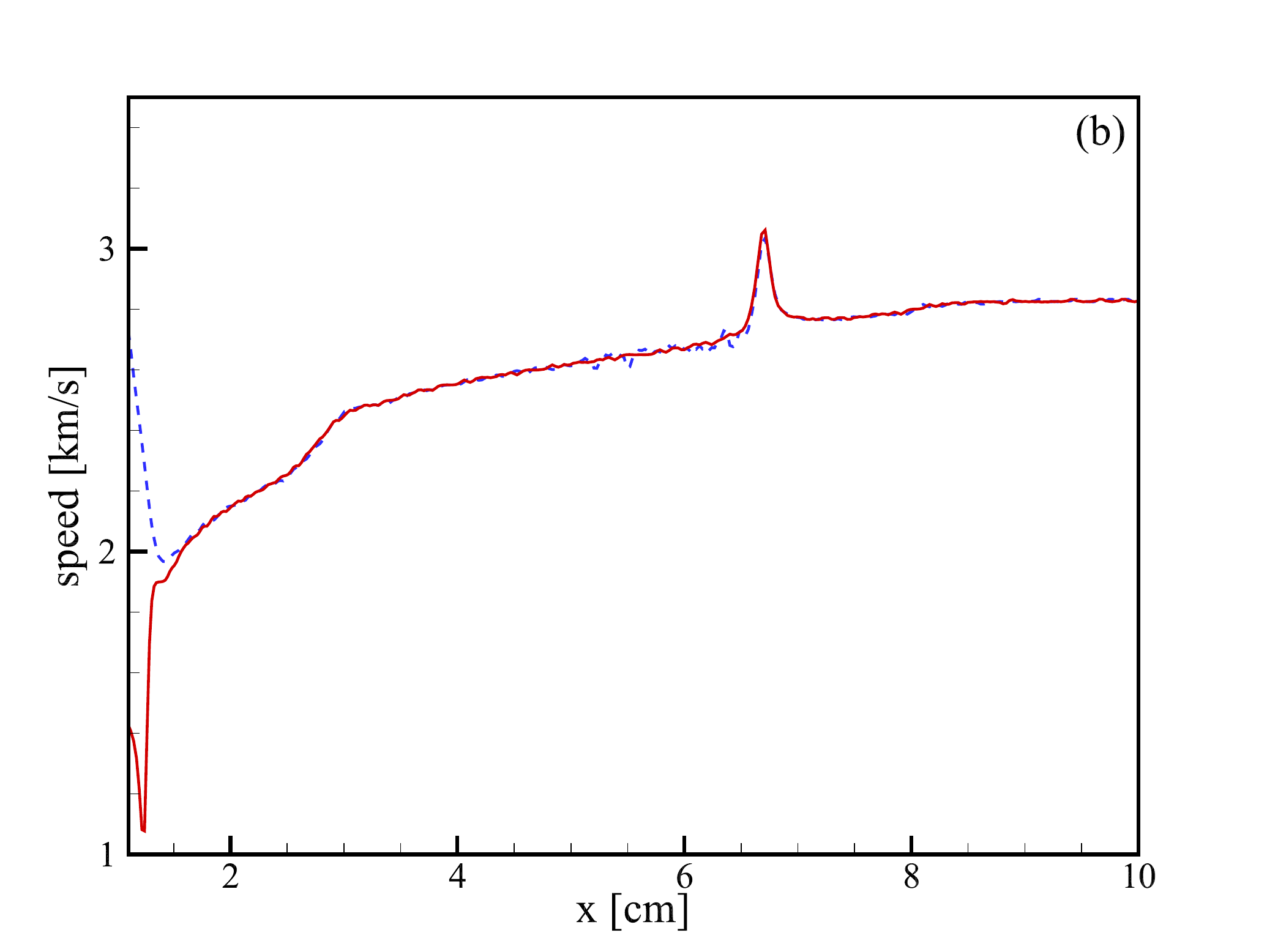}
\end{center}\caption[]{
Similar to \Fig{WENO_dx_10um}, but for $\delta x=5\um$ and with artificial
viscosity (colour online).
}\label{WENO_dx_5um_c_0-01}\end{figure*}

We use the artificial viscosity developed by \cite{KL2012} to increase the 
numerical dissipation of the WENO scheme. 
This does not contradict to the definition of convergence, because the 
artificial viscosity tends to zero, as $\delta x$ tends to zero. 
At the same resolution, however, the numerical dissipation of the WENO
scheme with artificial viscosity is larger than that of the WENO scheme.
The result with the WENO code with artificial viscosity for a resolution of 
$\delta x=5\mu m$ results in the development of a steady detonation, as shown 
in figure~\ref{WENO_dx_5um_c_0-01}. 
In simulations of problems containing shocks, we can calibrate the 
parameter of the artificial viscosity for a low resolution, but the problem of
detonation development (SWACER) is special, because in this case the 
parameter of artificial viscosity depends on the resolution, which makes 
the simulations much more demanding and time consuming, especially when we use 
detailed chemical models. 

It should be noted that the minimum resolution at which WENO code allows us to 
obtain a solution to the SWACER problem depends on the particular combustion 
gas mixture, the chemical kinetics scheme, and the initial pressure. 
At high resolution, the WENO code without artificial viscosity still shows 
the development of steady detonation for the mixture with high initial pressure. 
For example, for an initial pressure of 5~bar, a steady detonation develops
for the largest resolution of about $\delta x=2\um$ without the use of artificial 
viscosity.

\section{Conclusions}
\label{Concl}

Using high-resolution simulations of detonation initiated by an initial
temperature gradient in a hydrogen--oxygen mixture, we have shown,
using the {\sc Pencil Code}, that the
transition to and properties of detonation can successfully be modelled
for intermediate values of the shock viscosity parameter.
The numerical error, as determined by comparing with an empirical fit
to the pressure peak in the final stage of TD, is found to decrease
like $\delta x^{1.4}$ with decreasing mesh size
down to $\delta=0.2\um$.
(The typical performance is $0.05\us$ wall clock time per step and mesh
point with 2048 processors on a Cray XC40 with $2.3\GHz$ Intel cores.)
The shock viscosity has non-vanishing values only in the immediate
proximity of the shock and reaches there still values of about four
times the molecular value in our highest resolution simulation.
Unfortunately, the position of the shock still depends on the value
of $C_{\rm shock}$ of around $3\km\s^{-1}$.
Nevertheless, the shock speed reaches the expected value in the final
stage of TD.

It remains unsatisfactory that even at the largest resolution of half
a million mesh points in just the $x$ direction, we are still unable to
avoid the use of a shock viscosity.
This is because the shock is so strong and the molecular viscosity
still too small by comparison.
Furthermore, we have been unable to demonstrate that the use of a small
amount of shock viscosity does not affect the details of the shock
position or even the detailed shape of the shock profile.
We can therefore not be completely sure that TD will still
be recovered at even higher resolution, which has not yet been possible
to simulate.
A reason for the current limitation is that our code is optimised to
work for three-dimensional problems.
It is therefore conceivable that a significant speed-up could be achieved
by optimising the code for one-dimensional problems.
In that case it would also be rather straightforward to use an adaptive
mesh, which could make the calculations significantly more economic.

Another possible avenue for future research is to solve the governing
equations in conservative form so that mass, momentum, and energy
are conserved to machine precision.
A difficulty here is the presence of source terms in the equations for
the mass fractions of the individual species.

The WENO scheme is computationally demanding and it is
difficult to reach resolutions comparable to what has been done with
the {\sc Pencil Code}.
Nevertheless, a steady detonation front was obtained at the resolution 
of $\delta x=10\um$, and with the use of artificial viscosity at the 
resolution $\delta x=5\mu m$.
On the other hand, of course, we know from experiments that TD does
occur.
Thus, assuming that our equations are physically correct, as stated,
there should be no doubt that any failure to recover TD must be regarded
as a numerical artifact.

Yet another approach is to isolate the essence of the problem in a
simpler single reaction model.
One must then also use an idealised viscosity and a
simplified equation of state.
Those modifications could enable us to perform simulations at much higher
resolution so that it is possible to focus on the purely numerical aspect
of using a shock viscosity in this problem.

\section*{Acknowledgements}

This research was supported in part by the
National Key R\&D Program of China (Grant No.2018YFC0807900), 
the National Natural Science Foundation of China (grant 11732003),
the Beijing Natural Science Foundation (grant 8182050),
the National Science Foundation under the Astronomy and Astrophysics
Grants Program (grant 1615100),
and the University of Colorado through its support of the George Ellery
Hale visiting faculty appointment.
Simulations presented in this work have been performed with computing
resources provided by the Swedish National Allocations Committee at
the Center for Parallel Computers at the Royal Institute of Technology
in Stockholm.

\markboth{\rm C. QIAN ET AL.}{\rm GEOPHYSICAL \& ASTROPHYSICAL FLUID DYNAMICS}
\bibliographystyle{gGAF.bst}
\markboth{\rm C. QIAN ET AL.}{\rm GEOPHYSICAL \& ASTROPHYSICAL FLUID DYNAMICS}
\bibliography{paper}
\markboth{\rm C. QIAN ET AL.}{\rm GEOPHYSICAL \& ASTROPHYSICAL FLUID DYNAMICS}

\label{lastpage}
\end{document}